\documentclass[useAMS,usenatbib]{mn2e}

\usepackage{graphicx}
\usepackage{bm}
\usepackage{float}
\usepackage{multicol}
\usepackage{lscape}
\newcommand{\degree}{\ensuremath{^\circ}}
\newcommand\nodata{ ~$\cdots$~ }
\newcommand\ion[2]{#1$\;${\small\rmfamily\@{#2}}\relax}%

\def\aap{A\&A}
\def\apj{ApJ}
\def\aj{AJ}
\def\mnras{MNRAS}
\def\pasp{PASP}
\def\araa{ARA\&A}
\def\apjs{ApJS}
\def\procspie{Proc.~SPIE}
\def\aaps{A\&AS}
\def\apjl{ApJ}
\def\actaa{Acta Astron.}

\title[V356 Sgr's Environment]{The Complex Circumstellar and Circumbinary Environment of V356 Sgr}

\author[J.R. Lomax et al.]{Jamie R. Lomax$^{1,2}$\thanks{E-mail: jrlomax@uw.edu}, Andrew G. Fullard$^{3}$, Michael A. Malatesta$^{1}$, Brian Babler$^{6}$,\newauthor Daniel Bednarski$^{7}$, Jodi R. Berdis$^{1}$, Karen S. Bjorkman$^{4}$, Jon E. Bjorkman$^{4}$,\newauthor Alex C. Carciofi$^{7}$, James W. Davidson Jr.$^{5}$, Marcus Keil$^{1}$,  Marilyn R. Meade$^{6}$,\newauthor Kenneth Nordsieck$^{6}$, Matt Scheffler$^{1}$, Jennifer L. Hoffman$^{3}$, and John P. Wisniewski$^{1}$\\
$^{1}$Homer L. Dodge Department of Physics \& Astronomy, University of Oklahoma, 440 W Brooks Street, Norman, OK 73019, USA\\
$^{2}$Department of Astronomy, Box 351580, University of Washington, Seattle, WA 98195, USA\\
$^{3}$Department of Physics and Astronomy, University of Denver, 2112 East Wesley Ave., Denver, CO 80208, USA\\
$^{4}$Department of Physics \& Astronomy, University of Toledo, 2801 W Bancroft Street, Toledo, OH 43606, USA\\
$^{5}$Department of Astronomy, University of Virginia, P.O. Box 400325, Charlottesville, VA 22904-4325, USA\\
$^{6}$Department of Astronomy, University of Wisconsin-Madison, 475 N. Charter St., Madison, WI 53706, USA\\
$^{7}$Instituto de Astronomia, Geof\'{i}sica e Ci\^{e}ncias Atmosf\'{e}ricas, Universidade de S\~{a}o Paulo, Rua do Mat\~{a}o 1226, Cidade Universit\'{a}ria,\\ 05508-900 S\~{a}o Paulo, SP, Brazil}
\begin{document}

\date{Accepted . Received ; in original form }

\pagerange{\pageref{firstpage}--\pageref{lastpage}} \pubyear{2014}

\maketitle

\label{firstpage}

\begin{abstract}
We analyze 45 spectropolarimetric observations of the eclipsing, interacting binary star V356 Sgr, obtained over a period of $\sim$21 years, to characterize the geometry of the system's circumstellar material.  After removing interstellar polarization from these data, we find the system exhibits a large intrinsic polarization signature arising from electron scattering. In addition, the lack of repeatable eclipses in the polarization phase curves indicates the presence of a substantial pool of scatterers not occulted by either star. We suggest that these scatterers form either a circumbinary disk coplanar with the gainer's accretion disk or an elongated structure perpendicular to the orbital plane of V356 Sgr, possibly formed by bipolar outflows. We also observe small-scale, cycle-to-cycle variations in the magnitude of intrinsic polarization at individual phases, which we interpret as evidence of variability in the amount of scattering material present within and around the system. This may indicate a mass transfer or mass loss rate that varies on the time-scale of the system's orbital period. Finally, we compare the basic polarimetric properties of V356 Sgr with those of the well studied $\beta$ Lyr system; the significant differences observed between the two systems suggests diversity in the basic circumstellar geometry of Roche-lobe overflow systems.
\end{abstract}


\section{Introduction}
Single star evolution is characterized primarily by a star's initial mass as it reaches the zero-age main sequence. However, the evolutionary pathway of a star in a binary or multiple system is significantly harder to predict, because these stars can lose a large fraction of their mass through interactions with their companions. Companion-affected mass loss is particularly important for massive stars, which are predominantly found in binary systems; approximately 70\% of all O stars will have their evolution strongly affected by a companion \citep{sana2012}. 

One way that the stars in a massive binary system can interact is via Roche-lobe overflow \citep[RLOF;][]{sana2012}, during which a large fraction of the mass in a star's outer layers may be stripped and transferred to its companion \citep{Iben1991,Podsiadlowski1992,sana2012}. Despite the far-reaching effects of this process, it is not well understood. For example, the fractions of material transferred and lost from massive binaries undergoing RLOF are unknown \citep{deMink2007}. Because changes in mass loss rates as small as a factor of two can alter the evolutionary model tracks of massive stars enough to produce different types of supernovae than otherwise expected \citep[SNe;][]{Smartt2009}, quantifying mass loss in interacting binary systems is vital to understanding massive stellar evolution.

Unfortunately, directly imaging RLOF systems to better understand their mass-loss geometries remains largely beyond our current technological abilities. While interferometric techniques have had some success at resolving the components of massive binary systems undergoing RLOF in recent years (e.g., \citealt{zhao2008}, who imaged the tidal distortion of the primary star in the $\beta$ Lyrae system), very few RLOF objects have been observed with this technique.

Spectropolarimetry provides a powerful way to diagnose the distribution of circumstellar material without the need to resolve system components. This technique is particularly advantageous for eclipsing binary systems because their eclipses provide strong geometrical constraints that aid in the interpretation of polarimetric data. In the best-studied case, the eclipsing RLOF system $\beta$ Lyr, \cite{hoffman1998} used position angle rotations in emission lines and across the Balmer jump to infer the existence of bipolar outflows, independently confirming their interferometric detection by \cite{harmanec1996}. Similarly, \cite{lomax2012} used polarimetric light curves of the $\beta$ Lyr system to detect a hotspot on the edge of the disk, which reveals itself through a consistent offset between the timings of the secondary eclipse in total and polarized light. However, it remains unclear whether $\beta$ Lyr's circumstellar geometry is representative of most massive RLOF systems, or whether the system is unique.

V356 Sagittarii (hereafter V356 Sgr) is a massive interacting binary system whose geometry is thought to be similar to $\beta$ Lyr's. The 3 $M_{\odot}$, A2 supergiant secondary star  (hereafter ``donor'') has filled its Roche lobe and is currently transferring matter to the brighter, B3 V primary star \citep[hereafter ``gainer'';][]{wilson1978,peters2004,rensbergen2011}. The gainer, originally the less massive star, now has a mass of approximately 11 $M_{\odot}$ \citep{rensbergen2011}. A mass transfer stream between the two stars feeds an accretion disk around the gainer \citep{wilson1978}. Because the system's far ultraviolet \textbf{(UV)} emission lines exhibit no change in shape, velocity, or strength during eclipses, \citet{peters2004} suggested that V356 Sgr may also possess bipolar outflows or other material outside the orbital plane of the system. 

Evolutionary modeling by \cite{ziolkowski1985} suggested that the mass transfer in V356 Sgr is not conservative, and that some material has been lost from the system. In fact, \citet{hall1981} found that they needed a quadratic term in their ephemeris to account for the system's small change in period with time. However, many studies of the system have found a fairly stable 8.89-day period (e.g. Popper 1980). \citet{wilson1995} argued that mass transfer in the system must be intermittent to explain the lack of observed period change in some datasets. In fact, the quadratic term in \citet{hall1981}'s ephemeris produces only a 0.22 day (2.5\%) increase in period from its initial epoch to the date of the last measurement we present in this paper.

\citet{polidan} first reported that V356 Sgr exhibited a polarization signal. These authors showed that the system is intrinsically polarized at UV wavelengths, although they were unable to establish if the system is polarimetrically variable because there was only one observation. Because the two stars in the system are hot, the primary polarization mechanism in V356 Sgr is most likely electron scattering in its circumstellar material. Electron scattering is a wavelength-independent process that preserves information about the geometry of the scattering medium within the polarization signal of the scattered light. Therefore, analyzing the polarimetric behavior of V356 Sgr allows us to investigate the geometric properties of its circumstellar material without the need for interferometric techniques to resolve the system. In addition, the eclipsing nature of V356 Sgr provides constraints on the interpretation of the spectropolarimetric data, offering a unique opportunity to investigate the three-dimensional structure of its circumstellar material.

In this paper, we present the results of a broadband polarimetric study of V356 Sgr and compare its distribution of circumstellar material to that found in $\beta$ Lyr. In Section 2 we discuss the details of our observations. We present the results of our polarimetric data in Section 3, and interpret those results in the context of the geometry of the V356 Sgr system in Section 4. Finally, we summarize our findings in Section 5.

\section{Observational Data}
We monitored V356 Sgr at optical wavelengths over approximately 21 years using two different instruments. We also observed it once in the UV with a third facility. Below we describe the instrumental configurations we used and the resulting datasets.

\subsection{HPOL}
V356 Sgr was observed 29 times between 1994 and 1998 with the University of Wisconsin's Half-Wave Spectropolarimeter (HPOL) at the 0.9-m telescope at the Pine Bluff Observatory (PBO). In 2012 HPOL was moved to the 1-m telescope at the University of Toledo's Ritter Observatory (RO), where our team obtained an additional 12 observations. Observations before 1994 December 29 utilized a Reticon dual-channel photodiode array detector (hereafter Reticon) covering a wavelength range of 3200--7600 \AA{} with a spectral resolution of $15$ \AA{} \citep{wolff1996}.  After 1995 January 24, all data were recorded with a CCD detector using a blue (3200--6000 \AA{}) grating with a  resolution of 7.5 \AA{} and a red grating (6000--10,500 \AA{}) with a spectral resolution of 10 \AA{} \citep{nordsieck1996,davidson2014}. 

Our observations lasted approximately one hour for data obtained with the Reticon detector and one hour per grating with the CCD detector.  While we aimed to cover the entire wavelength range every night we observed with the CCD, poor weather and limited visibility of the system sometimes made this impossible. As a result, 9 of our 41 CCD observations used only a single grating (red or blue). We performed all basic data reduction steps using \texttt{REDUCE}, a custom software package developed for HPOL at the University of Wisconsin \citep{nook1990,wolff1996,davidson2014}.

Table \ref{tbl:hpoldata} lists the phases, civil dates, and Julian dates of our 41 observations.  We calculated the orbital phase for each observation by using the Julian date of the midpoint of each observation and the ephemeris reported by \citet{hall1981}. When weather conditions permitted both gratings to be used back to back, we used the start of the second observation as the midpoint Julian date. For all other observations, we calculated the midpoint of the observation by adding half the total exposure time to the starting Julian date. 

HPOL's instrumental polarization is updated on a quasi-monthly basis by observing known polarized and unpolarized standard stars. \citet{davidson2014} tabulated these values from 1995 January through 2013 May. Those authors also demonstrated that the instrumental polarization of HPOL in its new location at RO is comparable to, and often lower than, the instrumental polarization achieved when it was mounted at PBO. We calculated the root\textbf{-}mean\textbf{-}square systematic uncertainty for each CCD observation at RO from the individual \textit{q} and \textit{u} systematic uncertainties listed in \cite{davidson2014} in order to match the values historically reported for HPOL; we tabulate the results in Table \ref{tbl:hpoldata}. The Reticon systematic uncertainties are less well-known. However, our previous experience with HPOL data suggests that they are approximately $\boldsymbol{<}\space 0.02\%$. 

\subsection{IAGPOL}
We obtained additional observations of V356 Sgr on three nights in 2015 at the Observat\'{o}rio Pico dos Dias -- Brazil (OPD/LNA) using the 0.6\textbf{-}m Boller \& Chivens (B\& C) and the 1.6\textbf{-}m Perkin-Elmer (P-E) telescopes with the IAGPOL imaging polarimeter \citep{magalhaes1996}. IAGPOL consists of a calcite prism and a rotating half-wave plate, which is typically rotated through eight consecutive positions that are 22.5$^{\degree}$ apart to produce one polarimetric observation per \textit{BVR} or \textit{I} broadband filter. A sequence of images are taken at each waveplate position and are later summed during the reduction process. At least one standard star is observed each night to calibrate the polarization position angle. Additional details concerning the observational setup and the procedure used to reduce our IAGPOL data are given in \citet{carciofi2007} and references therein.

Table \ref{tbl:hpoldata} lists the phases, civil dates, and Julian dates of our IAGPOL observations. Our listed Julian dates are the mean Julian date of the observation, which was computed by taking average of the Julian dates of the first exposure taken at the first waveplate position and the last exposure taken at the last waveplate position. Phases were then calculated using the \cite{hall1981} ephemeris. Each night we observed the V356 Sgr system using IAGPOL's full filter set, except on 29 April 2015, when we obtained a second observation with the \textit{B} filter. This observation was started immediately after the completion of an observation using the full filter set. Therefore, we list it as a separate observation. Depending on the night and the filter either 8 or 16 waveplate positions were used. Exposure times ranged from 0.21 s to 1.5 s and we typically obtained 30 exposures per waveplate position, except on the night of 28 July 2015 when we obtained 35 exposures per waveplate position. Previous experience with IAGPOL indicates that its observational setup produces an instrumental polarization of $\boldsymbol{<}\space 0.005\%$ on each telescope. 

\subsection{WUPPE}
V356 Sgr was observed once by the Wisconsin Ultraviolet Photo-Polarimeter Experiment (WUPPE) as part of the ASTRO-2 mission aboard the Space Shuttle Endeavour. WUPPE was a 0.5-m f/10 Cassegrain telescope and UV spectropolarimeter that obtained data between 1,400 and 3,200 \AA\space at a spectral resolution of 16 \AA. Further WUPPE information can be found in \cite{Bjorkman} and \cite{wuppe}. V356 Sgr was observed on 05 March 1995 with a 932 s exposure time. The observation's midpoint Julian date is $2,449,781.857$ and it has a phase of 0.149.

\begin{table*}
\begin{tabular}{lcccccccccc}
\hline
Julian Date & Civil Date & Filter & \% q & \% u & Int. Error  & Sys. Error & Pol.  & P.A.  & P.A. Error  & Phase \\
& & & & & (\%) & (\%) & (\%) & (\degree) & (\degree) & \\
\hline
\hline
\multicolumn{11}{c}{HPOL, Reticon at PBO}\\
\hline

 2449624.12 &   1994 Sept 29 & B & -0.071 & 0.785 & 0.027 & 0.020 & 0.789 & 47.59 & 0.99 & 0.430 \\  
\nodata  & \nodata  & V & -0.210 & 0.878 & 0.025 & 0.020 & 0.903 & 51.72 & 0.79 &  \nodata   \\    
\nodata  & \nodata  & R & -0.159 & 0.866 & 0.032 & 0.020 & 0.880 & 50.19 & 1.04 &  \nodata   \\    

2449625.09 &   1994 Sept 30 & B & -0.308 & 0.712 & 0.022 & 0.020 & 0.776 & 56.68 & 0.80 & 0.539 \\  
\nodata  & \nodata  & V & -0.300 & 0.842 & 0.018 & 0.020 & 0.894 & 54.79 & 0.56 &   \nodata  \\    
\nodata  & \nodata  & R & -0.353 & 0.939 & 0.022 & 0.020 & 1.003 & 55.31 & 0.62 &  \nodata   \\    

\hline
\multicolumn{11}{c}{HPOL, CCD at PBO}\\
\hline

2449856.91 &   1995 May 19 & B & -0.244 & 0.894 & 0.016 & 0.010 & 0.926 & 52.63 & 0.49 &  0.598  \\  
\nodata  & \nodata  & V & -0.295 & 0.759 & 0.007 & 0.005 & 0.814 & 55.62 & 0.25 &   \nodata \\    
\nodata  & \nodata  & R & -0.268 & 0.585 & 0.004 & 0.004 & 0.643 & 57.30 & 0.17 &   \nodata \\    
\nodata  & \nodata  & I & -0.400 & 0.760 & 0.005 & 0.007 & 0.858 & 58.88 & 0.18 &   \nodata \\

\end{tabular}
\caption{Total (intrinsic plus interstellar) optical polarization for each of our spectropolarimetric observations, convolved with synthetic Johnson-Cousins broadband filters. The quoted phases are calculated from the ephemeris given by \citep{hall1981}. ``PBO'' denotes data obtained with the HPOL instrument mounted at Pine Bluff Observatory, while ``RO'' denotes data obtained with the HPOL instrument mounted at Ritter Observatory ($\S$ 2). ``B\&C'' denotes data taken with the IAGPOL instrument on the Boller \& Chivens 0.6m telescope, and ``P-E'' indicates data taken with IAGPOL on the Perkins-Elmer 1.6m telescope at the Observat\'{o}rio Pico dos Dias. The full table is available online.}
\label{tbl:hpoldata}
\end{table*}

\section{Results}

\subsection{Total Broadband Polarization}
 
We searched our optical polarimetric spectra for signatures of line polarization, but found none. Therefore, we do not display any optical polarimetric spectra of V356 Sgr. In order to characterize the behavior of the system, we convolved each of our HPOL spectropolarimetric observations with synthetic broadband Johnson-Cousins \citep [\textit{BVRI};][]{bessell1990} filter equivalents. In Table \ref{tbl:hpoldata}, we tabulate these values, their associated internal uncertainties based on photon-noise statistics, and their systematic instrumental uncertainties (see Section 2.1; \citealt{davidson2014}) for each observation when its spectrum covers the full wavelength range of our synthetic filters. Table 1 also lists our broadband IAGPOL \textit{BVRI} data. In all of our figures, the displayed HPOL uncertainties represent the larger of the observations' internal and systematic errors. Because IAGPOL uncertainties are dominated by photon noise, our displayed error bars on those points represent the internal error only.

We find that in all optical photometric bands, the polarization remains approximately constant as a function of phase. The data also appear to exhibit prominent non-periodic, cycle-to-cycle stochastic variations, which produce large scatter at many individual phases. The amplitude of the stochastic variability is approximately 0.25--0.50\%, which is substantially larger than both the formal and systematic uncertainties of these data. The polarization in each band arises from two sources: scattering in the circumstellar material around V356 Sgr (i.e. the intrinsic polarization component) and dichroic absorption in the interstellar medium located along the line of sight to V356 Sgr (i.e. interstellar polarization or ISP). Because we expect the interstellar polarization is constant over periods longer than the time-scale over which our data were taken, we consider this behavior to be intrinsic to the system. If changes in the scattering angle due to the system's orbit were the only cause of variations, we would expect the polarization to return to the same value at the same phase for each cycle \citep{brown1978}. Because that does not occur, it is clear that variations in the electron density of the circumstellar material are the dominant cause of the observed stochastic variations.

\begin{figure}
\centering
\includegraphics[width=0.48\textwidth]{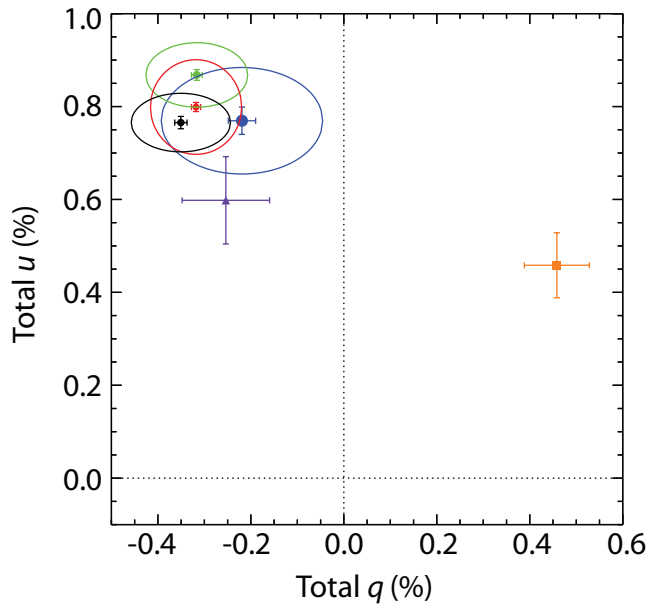}
\caption{The Stokes \textit{q-u} plot of the mean total (intrinsic plus interstellar) optical polarization of V356 Sgr in the \textit{B} (blue filled circle, filled circle in black and white text version), \textit{V} (green filled circle, filled diamond in black and white text version), \textit{R} (red filled circle, asterisk in black and white text version), and \textit{I} (black filled circle, filled downward triangle in black and white text version) bands. The point size represents the error on the mean polarization. Error bars on these points represent the average observational error in each band while the ellipses represent the RMSD from the mean polarization in each band. The purple filled triangle and its error bars (filled upward triangle in black and white text version) represent the error-weighted mean UV polarization from our WUPPE spectrum and its associated uncertainty. The orange filled square represents our ISP estimate (Section 3.2) at $\lambda_{max}$ and its associated uncertainty.}
\label{qu}
\end{figure}

In Figure \ref{qu}, we display a Stokes \textit{q-u} plot of the mean polarization in each of the optical bands. To characterize the large amount of stochastic scatter in our data, we calculate the average root-mean-square deviation (RMSD) from the mean in both \textit{q} and \textit{u} for each observation in each band. We display the RMSD in \textit{q} and \textit{u} as ellipses centered on the mean broadband polarization in Figure \ref{qu}. The polarization in all bands is consistent within uncertainties. This suggests that the polarization and position angle are relatively constant with wavelength. 

In order to characterize the broadband UV behavior of V356 Sgr, we also calculated the error-weighted mean \textit{q} and \textit{u} for the entire WUPPE spectrum (i.e. no synthetic filters were used). We display this in Figure \ref{qu}. In this case, our error bars represent the traditional error on the mean. As can be seen from the \textit{q-u} plot, the position angle of the UV data is consistent with the average position angle of the optical data, but the magnitude of the polarization appears to be slightly smaller in the UV. In order to more robustly characterize the system's polarization and assess its origin, we isolate the intrinsic components in both the optical and UV below. 

\subsection{Interstellar Polarization}

Accurately characterizing and removing interstellar polarization (ISP) along the line of sight to V356 Sgr is a challenging yet necessary task that needs to be completed before we can investigate the origin of V356 Sgr's intrinsic polarization component. While numerous techniques to determine ISP have been detailed in the literature \citep{mcl79, qui97, cla98, wis10, dra14}, the optimal technique must generally be determined on a case-by-case basis.

We used the field-star method, whereby one characterizes the ISP by averaging the observed polarization of stars physically close to the object of interest (i.e. stars that are a similar distance from Earth and that have small angular separations from the object). This technique assumes that the interstellar medium is effectively homogeneous over small spatial scales and that each of the field stars used is itself devoid of an intrinsic polarization component. The field-star method works best when determining the ISP in the direction of a star cluster, in which all stars are spatially nearby one another and located at approximately the same distance \citep{wis07}.  

\begin{figure}
\centering
\includegraphics[width=0.48\textwidth]{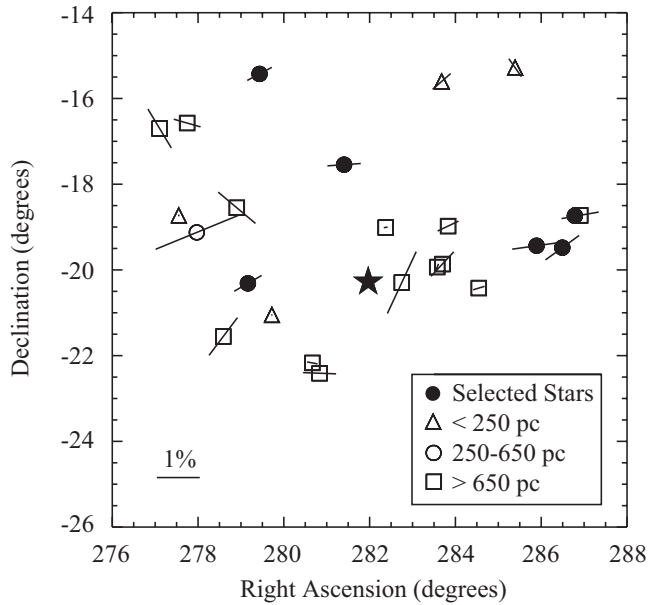}
\caption{\textit{V}-band polarization of field stars compiled in Table \ref{tbl:fieldstars} as a function of their spatial separation from V356 Sgr.  Filled circles indicate the set of stars used for the weighted field star average, whereas open squares denote field stars that were omitted from the average as discussed in $\S$ 3.2.  The filled star in the center of the figure represents V356 Sgr.}
\label{fig:fieldstars}
\end{figure}

We compiled a list of candidate field stars located within six degrees of V356 Sgr from the \cite{Hall1958}, \cite{mat70}, and \cite{heiles2000} catalogs. Objects classified in the SIMBAD catalog \citep{SIMBAD} as known Be stars were immediately cut from our list because they are known to be intrinsically polarized (e.g., \citealt{wis10}). Table \ref{tbl:fieldstars} lists our remaining field stars; we display their locations on the sky in Figure \ref{fig:fieldstars}. V356 Sgr is located at a distance of $454\pm212$ pc \citep{vanleeuwen2007}; therefore, we narrowed our list of viable field stars to those located between 250 and 650 pc (within the $1\sigma$ uncertainty of V356 Sgr's distance). Next we removed stars that could plausibly exhibit some intrinsic polarization, specifically those identified as variable stars in the SIMBAD catalog. For example, HD 170764 meets our distance and small angular separation criteria, but we excluded it because it is a known Cepheid variable star \citep{egg57}. Of the 7 remaining stars, we excluded HD 177863, which is categorized as a spectroscopic binary in SIMBAD, because the position angle of its polarization (135.8$\pm 2.8\degree$) is significantly different than the position angles of the other 6 remaining field stars (average 18.9$\pm 2.8\degree$). We adopted the remaining 6 stars, labeled with ``a'' in Table 2 and shown as filled circles in Figure \ref{fig:fieldstars}, as our final group of field stars. We note that one of these 6 stars (HD 177559) is classified as an eclipsing binary in SIMBAD, but we retained it because its polarization is consistent with the other remaining stars and, given the ubiquity of binary systems, it is likely that one of our other field stars is an unknown binary. We computed the error-weighted mean polarization and position angle quoted in Table \ref{tbl:fieldstars} from the error-weighted average $q$ and $u$ of the 6 final sources. We also calculated a distance-weighted average, but the distance-weighted average polarization was the same as the error-weighted mean polarization. While the distance-weighted mean position angle was 10\degree\space greater than the position angle calculated from the error-weighted mean, it did not significantly change the intrinsic polarization and position angles we calculated after removing the ISP (see next paragraph); the intrinsic polarization and position angle of each observation was the same within uncertainties regardless of the weighting mechanism we used for the field stars. Therefore, we used the error-weighted average field star polarization and position angle in our ISP subtraction.

We determined the wavelength dependence of the ISP in the direction of V356 Sgr using the Wilking-modified Serkowski law, $P_\lambda = P_{max} \exp(-K \ln ^{2} (\lambda_{max}/ \lambda) )$, where $K=-0.1+1.86 \lambda_{max}$ and $\lambda_{max}$ is in microns \citep{serkowski1975,wil82}. We adopted the average polarization and position angle from the field-star method described above as our $P_{max}$ and ISP position angle. Because the three catalogs we used in our ISP estimate only report \textit{V}\textbf{-}band data, we could not use them to estimate $\lambda_{max}$. There were also no stars within 5 degrees of V356 Sgr with measured $\lambda_{max}$ values in any of the catalogs. Therefore, we took the value for $\lambda_{max}$ to be the median sky value from \cite{serkowski1975}, 5400 \AA. The final parameters for our ISP estimate are $P_{max}=0.58 \pm 0.07$ \%, $\lambda_{max}=5400$ \AA, $K=0.9044$, and $\theta=18.9\degree \pm 2.8$\degree. We display our ISP estimate in \textit{q-u} space in Figure \ref{qu}. We subtracted this estimate from our HPOL data in order to obtain ISP\textbf{-}removed polarimetric spectra for V356 Sgr. For the IAGPOL dataset, we calculated the expected ISP based on our Serkowski law parameters for the effective central wavelengths of the \textit{BVRI} filters (4361 \AA, 5448 \AA, 6407 \AA, and 7980 \AA\space respectively; \citealt{Bessell2005}), and subtracted those values from our observations. Any remaining polarization in our datasets should now be intrinsic to the system i.e.\textbf{,} caused by scattering in circumstellar material and not the interstellar medium. 

\begin{table*}
\begin{tabular}{lccccccc}
\hline
Field Star & Distance (pc) & Separation (\degree) & Pol. (\%) & Pol. Error (\%) & P.A. (\degree) & P.A. Error (\degree) & Ref.\\
\hline
\hline
V356 Sgr & 454.5 $\pm$ 212.8 & \nodata & \nodata & \nodata & \nodata & \nodata &\nodata \\
HD170433 &  83.2 & 4.44 & 0.04  & 0.035 & 96.2  & 23.6 & 2\\ 
HD170764 &  640  & 3.93 & 2.072 & 0.155 & 22.3  & 2.1 & 2\\ 
HD170978$^{a,b}$ &  382.5 & 5.12 & 0.269 & 0.121 & 4.8   & 12.7 & 1,2 \\
HD171611$^a$ &  316.2 & 2.63  &  0.74 & 0.2   & 30    & 7.7 & 2 \\
HD171876$^a$ &  380.2 & 5.41  &  0.65 & 0.2   & 28    & 8.7 & 1 \\ 
HD172051 &  15.1 & 2.24 &  0.03 & 0.035 & 23.4  & 30.3 & 2 \\  
HD172854 &  779  & 2.38 &  0.77 & 0.1   & 177.9 & 3.7 & 2 \\
HD173375$^a$ &  264 & 2.78  &  0.78 & 0.2   & 4     & 7.3 & 2 \\ 
HD174383 &  860 & 0.73 &  1.59 & 0.12  & 65    & 2.2 & 2\\ 
HD175043 &  882 & 1.55 &  0.16 & 0.1   & 67.7  & 17.4 & 2 \\
HD175141 &  724.4 & 1.68 &  0.78 & 0.2   & 47    & 7.3 & 2\\ 
HD175156 &  182 & 4.95 &  0.56 & 0.035 & 41.1  & 1.8 & 2\\ 
HD175253 &  722.5 & 2.18 &  0.54 & 0.1   & 25.5  & 5.3 & 2\\
HD176593 &  100 & 5.96  &  0.51 & 0.035 & 125   & 2  & 3\\ 
HD177014$^a$ &  575.8 & 3.79 &  1.15 & 0.2   & 8     & 5 & 2 \\ 
HD177559$^a$ &  501.2 & 4.33 &  0.97 & 0.2   & 36    & 5.9 & 2 \\ 
HD177863 &  266 & 4.79  &  0.36 & 0.035 & 135.8  & 2.8 & 2\\
Adopted Field Star Mean & \nodata &\nodata & 0.58 & 0.07 & 18.9 & 2.8 &\nodata \\
\hline
\end{tabular}
\caption{Published \textit{V}-band polarization of field stars within a $6\degree$ angular separation of V356 Sgr after an initial cut to remove known Be stars and stars whose distance is greater than 1000 pc. We compiled this list from the (1) \citealt{Hall1958}, (2) \citealt{heiles2000}, and (3) \citealt{mat70} catalogs. $^a$We used these stars, located within the $1\sigma$ uncertainty of the distance to V356 Sgr (i.e., 250--650 pc; \citealt{vanleeuwen2007}) and having no suspected intrinsic polarization, for our final adopted field star polarization (Figure \ref{fig:fieldstars}). The error-weighted mean of the \textit{q} and \textit{u} values of these six stars yield the ``Field Star Average'' values quoted in the last row. $^b$Values quoted here represent the error-weighted mean polarization and position angle values calculated by finding the mean \textit{q} and \textit{u} of} the two stated references.
\label{tbl:fieldstars}
\end{table*}

\subsection{Intrinsic Optical Polarization}

\begin{figure*}
\centering
\includegraphics[width=0.8\textwidth]{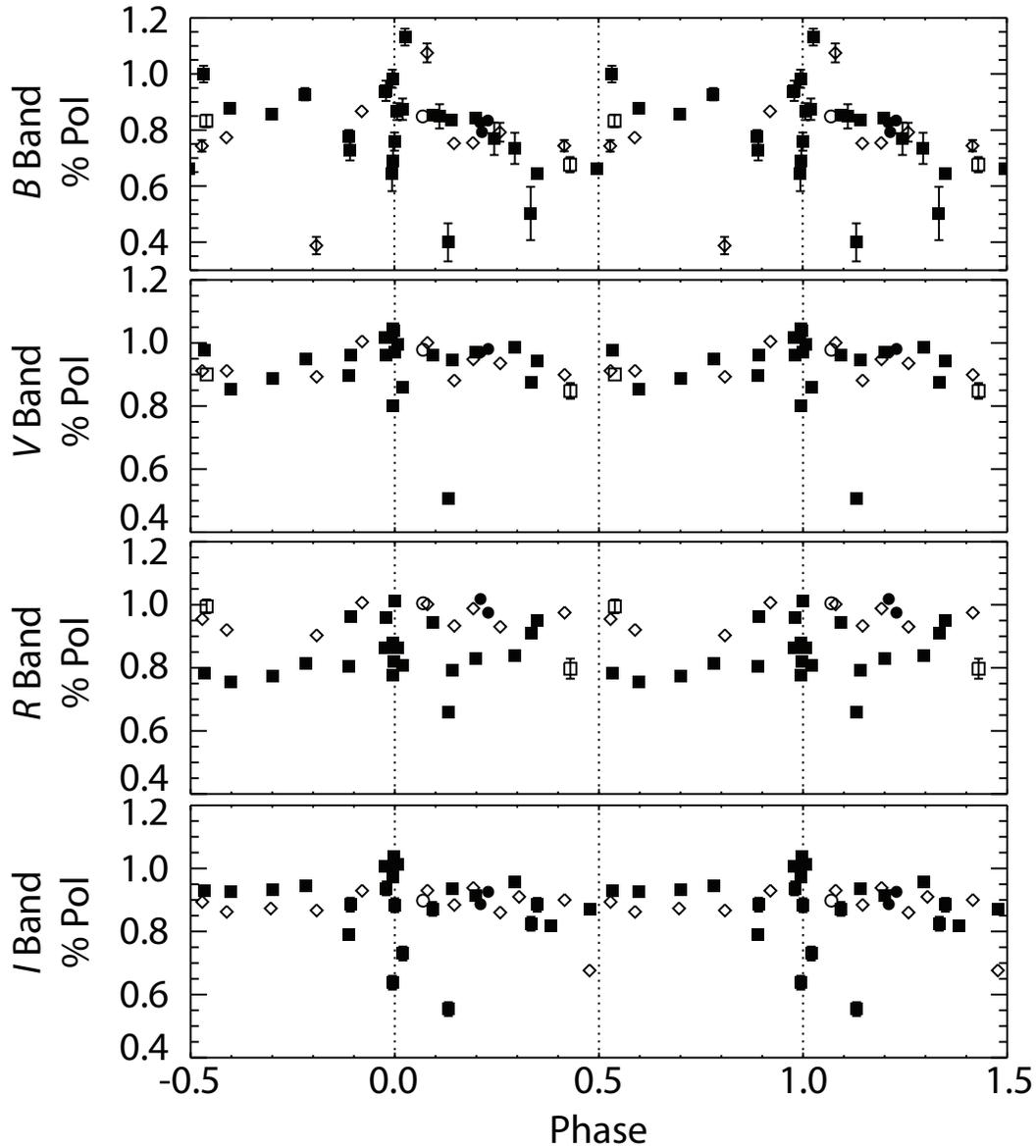}
\caption{Intrinsic (ISP-subtracted) optical polarization of V356 Sgr in the \textit{BVRI} bands as a function of orbital phase \citep{hall1981}. Open squares represent Reticon data taken at PBO, filled squares represent CCD data taken at PBO, open diamonds represent CCD data taken at RO, filled circles represent IAGPOL data taken at the B$\&$C telescope, and open circles represent IAGPOL data taken at the P-E telescope (Section 3.3; Table \ref{tbl:intrins}). Data are folded in phase and wrapped so that more than one cycle is shown. Uncertainties smaller than our point size are not displayed. The locations of photometric eclipses are marked by dashed, vertical lines. Phases 0.0 and 1.0 represent primary eclipse (disk and gainer eclipsed) while phase 0.5 represents secondary eclipse (donor eclipsed). The polarization shows no clear phase-dependent behavior in any of the filters, although prominent epoch-to-epoch variations for a given phase are clearly present.}
\label{fig:phasepolsub}
\end{figure*}

Similar to our analysis in Section 3.1, we applied synthetic \textit{BVRI} Johnson-Cousins filter equivalents \citep{bessell1990} to our intrinsic (ISP-removed) polarized HPOL spectra. Table \ref{tbl:intrins} lists our HPOL and IAGPOL intrinsic broadband polarization data and their internal uncertainties. Because the systematic uncertainty for each observation is not affected by our ISP removal process, we do not re-tabulate those values. Figure \ref{fig:phasepolsub} displays these data graphically. The error bars displayed in Figure \ref{fig:phasepolsub} represent the larger of the internal and systematic uncertainties and therefore do not always correspond to the values tabulated in Table \ref{tbl:intrins}.

After we subtracted our ISP estimate, the data still exhibit non-periodic, cycle-to-cycle stochastic variations. This is expected because we assume the ISP is constant with time over the period of our observations. The polarization in each band does not appear to vary with wavelength or phase. In order to characterize the average polarization and the amount of stochastic variability at various phases, we calculated both the error-weighted mean percent polarization and position angle in \textit{q-u} space, as well as the standard deviation of our data, in four phase bins corresponding to primary eclipse (phases 0.95-0.05), secondary eclipse (phases 0.45-0.55), out-of-eclipse phases (phases 0.05-0.45 and 0.55-0.95), and the full binary period (phases 0.0-1.0). We tabulate these results in Table \ref{tbl:avgintrins}.

\begin{figure*}
\centering
\includegraphics[width=0.7\textwidth]{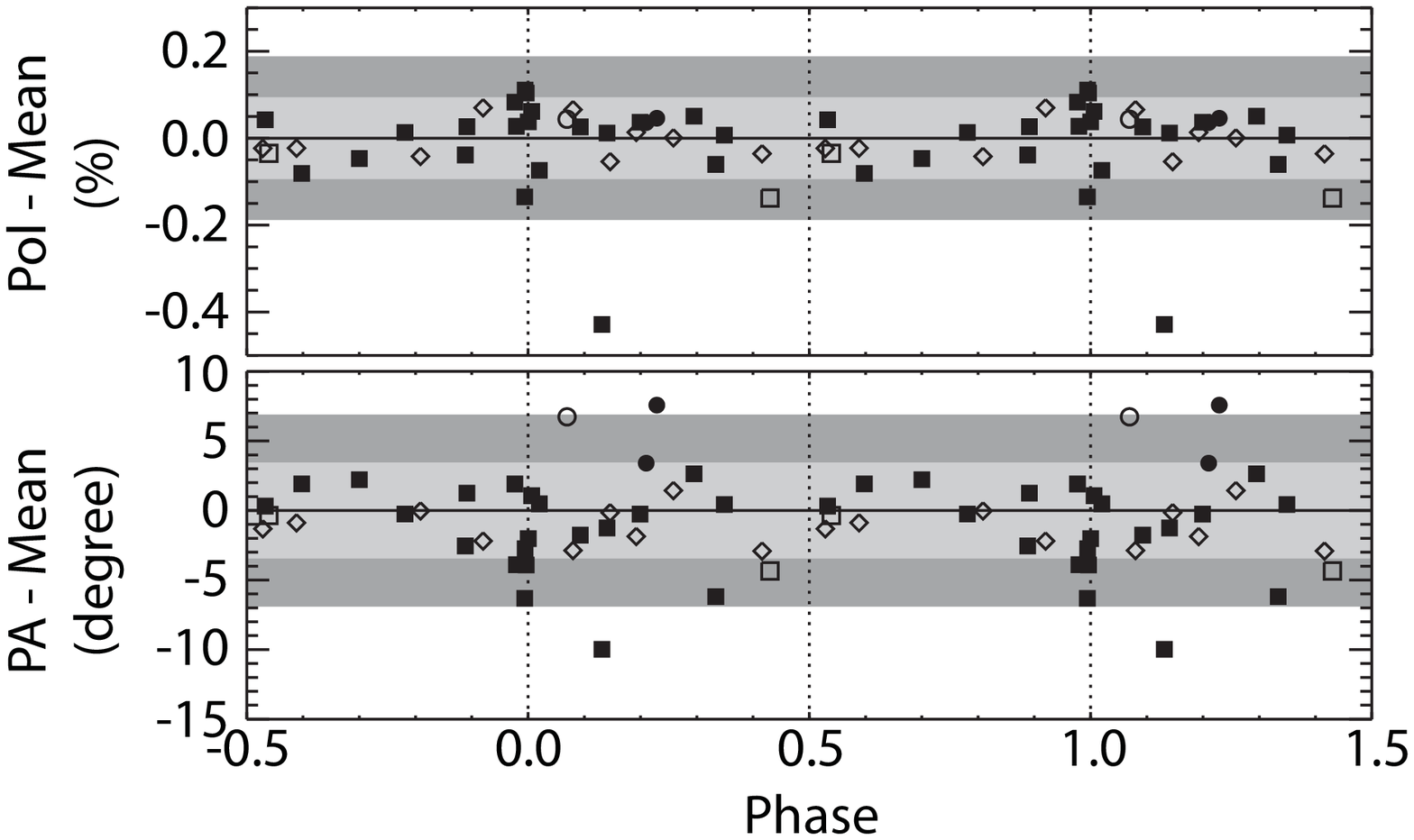}
\caption{Residuals derived from subtracting the mean global intrinsic \textit{V}-band polarization (top) and position angle (bottom) from each observation. The 1$\sigma$ and 2$\sigma$ standard deviations of the data are depicted by the light gray and dark gray boxes respectively. Point styles are the same as in Figure 3. Although the polarization near primary eclipse appears to have larger deviations from the mean than other phases, few individual observations exhibit a $>1$$\sigma$ variation. The position angles of our observations appear to deviate more from the mean, but most are still within 2$\sigma$. More significant deviations from the mean occur in the first half of the orbit than the second half. However, the second half of the orbit is less-well covered.}
\label{fig:sigma}
\end{figure*}

The error-weighted mean polarization is approximately constant ($\approx 0.8-0.9$ \%) across all phase bins (Table \ref{tbl:avgintrins}) and wavelengths. This shows that V356 Sgr's average polarimetric behavior is indeed non-periodic in nature and also consistent with electron scattering. Additionally, the standard deviation of the data is constant ($1\sigma \approx 0.1$ \%) with phase. We subtracted the mean global \textit{V}-band polarization (0.935\% ) in \textit{q-u} space from our dataset and display the residuals in Figure \ref{fig:sigma}. Most of the observed scatter is of low amplitude (below 2$\sigma$) and we find no convincing trend in the stochastic variations in our dataset as a function of phase. 

\begin{table*}
\begin{tabular}{lccccccc}
\hline
Julian Date & Civil Date & Filter & Pol. (\%) & Pol. Error (\%) & P.A. (\degree) & P.A. Error (\degree) & Phase\\
\hline
\hline

2449624.12 & 1994 Sep 29 & B & 0.676 & 0.027 & 69.33 & 1.16 & 0.430 \\
\nodata & \nodata & V & 0.848 & 0.025 & 70.91 & 0.84 & \nodata \\
\nodata & \nodata & R & 0.797 & 0.032 & 69.60 & 1.15 & \nodata \\
2449625.09 & 1994 Sep 30 & B & 0.833 & 0.022 & 76.68 & 0.75 & 0.539 \\
\nodata & \nodata & V & 0.900 & 0.020 & 73.59 & 0.56 & \nodata \\
\nodata & \nodata & R & 0.995 & 0.022 & 71.67 & 0.62 & \nodata \\
2449856.91 & 1995 May 19 & B & 0.878 & 0.016 & 70.46 & 0.51 & 0.598 \\
\nodata & \nodata & V & 0.854 & 0.007 & 75.86 & 0.24 & \nodata \\
\nodata & \nodata & R & 0.756 & 0.004 & 80.93 & 0.15 & \nodata \\
\nodata & \nodata & I & 0.926 & 0.007 & 75.97 & 0.17 & \nodata \\

\end{tabular}
\caption{We computed the intrinsic polarization of V356 Sgr by removing the interstellar polarization along the line of sight quoted in Section 3.2 and then applying synthetic broad-band \textit{BVRI} filters to the resulting spectra (Section 3.3). The full table is available online.}
\label{tbl:intrins}
\end{table*}

\begin{table*}
\begin{tabular}{lcccccccc}
\hline
Phase Range & B & $\sigma_B$ & V & $\sigma_V$ & R & $\sigma_R$ & I & $\sigma_I$\\
\hline
\hline
\multicolumn{9}{c}{Polarization (\%)}\\
\hline
0.95-0.05 & 0.911 & 0.150 & 0.971 & 0.083 & 0.861 & 0.075 & 0.978  & 0.155 \\
0.45-0.55 & 0.747 & 0.140 & 0.938 & 0.035 & 0.868 & 0.101 & 0.864 & 0.100 \\
0.05-0.45 \& 0.55-0.95 & 0.799 & 0.148 & 0.927 & 0.097 & 0.869 & 0.099 & 0.898 & 0.078  \\
0.0-1.0 & 0.804 & 0.149 & 0.935 & 0.092 & 0.867 & 0.094 & 0.905 & 0.100 \\
\hline
\multicolumn{9}{c}{Position Angle (deg)}\\
\hline
0.95-0.05 & 75.09 & 4.79 & 73.04 & 2.87 & 76.68 & 4.44 & 73.57 & 2.05 \\
0.45-0.55 & 74.99 & 1.65 & 73.44 & 0.67 & 75.29 & 3.06 & 75.83 & 1.96 \\
0.05-0.45 \& 0.55-0.95 & 76.21 & 8.13 & 74.20 & 3.68 & 76.30 & 4.07 & 75.26 & 3.53 \\
0.0-1.0 & 76.00 & 7.08 & 73.94 & 3.41 & 76.25 & 4.06 & 75.05 & 3.28\\
\hline
\end{tabular}
\caption{Error-weighted mean intrinsic polarization (\%\textit{p}) and position angle of V356 Sgr and their associated standard deviations for the phase bins described in Section 3.3. The mean intrinsic polarization and its standard deviation remain roughly constant across all phase bins and wavelength bands, as does the mean position angle. }
\label{tbl:avgintrins}
\end{table*}

\subsection{Optical Polarized Flux}
\begin{figure*}
\centering
\includegraphics[width=0.8\textwidth]{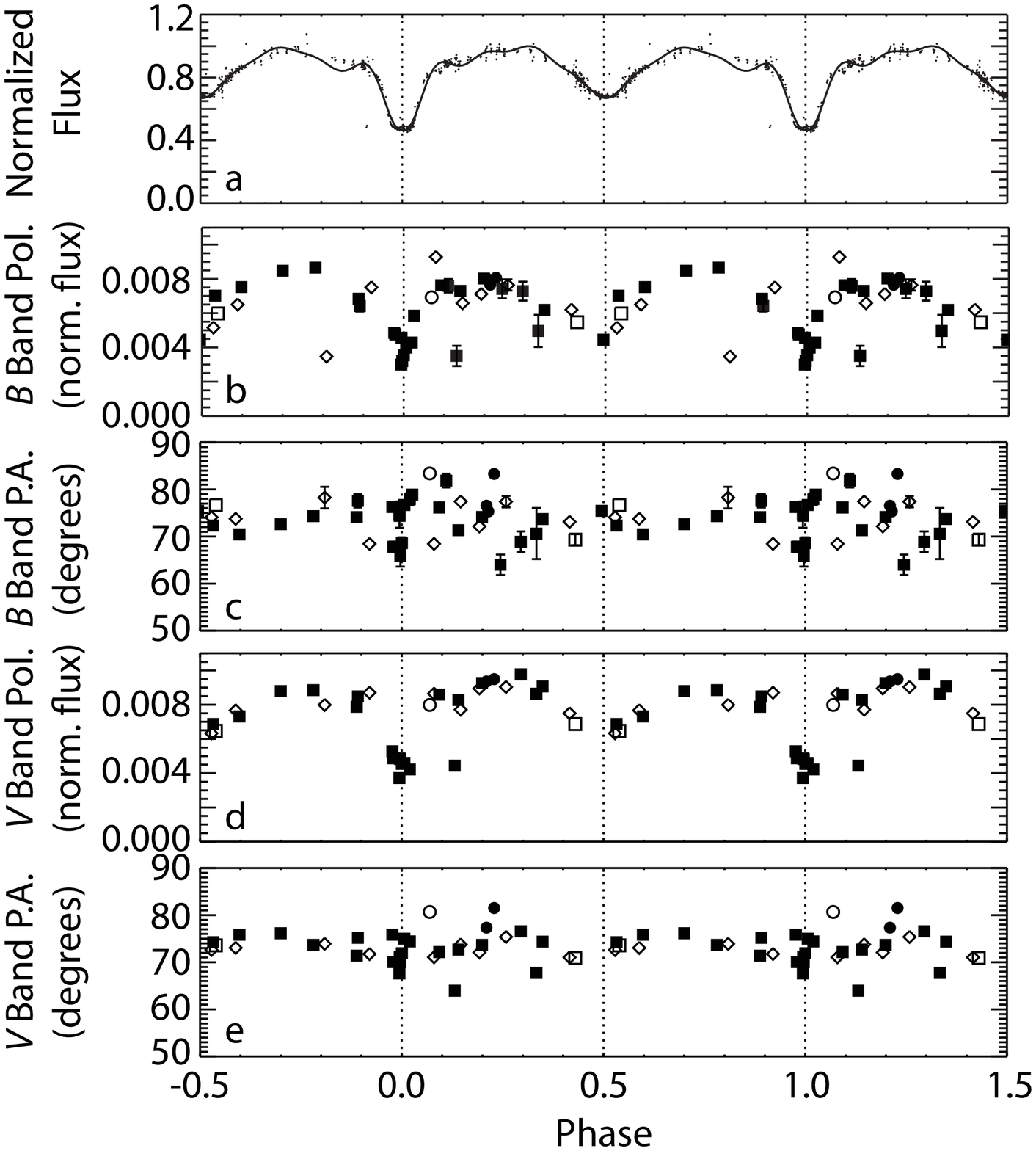}
\caption{\textit{From top:} \textbf{Phase} dependence of (a) the normalized \textit{y}-band photometry from \citet{popper1955} (small dots) and our Fourier fit to these data (solid line; Section 3.4); (b) \textit{B}-band relative polarized flux; (c) \textit{B}-band position angle; (d) \textit{V}-band relative polarized flux, and (e) \textit{V}-band position angle. Symbols have the same meanings as in Figure \ref{fig:phasepolsub}. Uncertainties are displayed only when they are larger than the point size. V356 Sgr's observed polarized flux is dominated by changes in the system's flux and not changes in its measured polarization.}
\label{fig:pflux}
\end{figure*}

Polarized flux light curves of systems can aid in the interpretation of the geometry of unresolved circumstellar environments because they can reveal the locations of the illumination sources as seen by the polarizing scatterer when compared to the percent polarization phase curves. Because HPOL and IAGPOL do not measure absolute fluxes, we created polarized flux light curves of V356 Sgr by performing a Fourier fit to the \textit{y}-band photometric data from \citet{popper1955} with \texttt{PERIOD04} \citep{lenz2005} and multiplying the resulting normalized curve by our \textit{B} and \textit{V} polarization band data. \texttt{PERIOD04} produces a Fourier fit of the form $y= Z + \sum_{i=1}^n A_i \sin(2\pi (\Omega_i t + \phi_i))$ where \textit{n} is the number of sine terms used, \textit{Z} is the zero point, \textit{A} is the amplitude, $\Omega$ is the frequency, and $\phi$ is the phase (note: this is not the same phase as the phase we calculated from V356 Sgr's ephemeris). Table \ref{tbl:fitparams} lists our Fourier fit parameters to the \textit{y}-band data and their associated uncertainties, which we determined using \texttt{PERIOD04}'s Monte Carlo uncertainty estimation algorithm. The top panel of Figure \ref{fig:pflux} displays our normalized Fourier fit \textit{y}-band light curve. The deviation of the fit from the \textit{y}-band data between phase 0.8 and 0.9 has only a minor impact on our polarized flux because our observational coverage of this part of the light curve is limited to 2 points in the \textit{B} and \textit{V} bands.

We used this y-band Fourier-fit light curve to calculate polarized fluxes in the \textit{B} and \textit{V} bands (Figure \ref{fig:pflux}, panels b and d). The variation in the polarized flux with phase mirrors that of the total flux, with distinct eclipses at phases 0 and 0.5. We discuss the implications of this result in Section 4.

\begin{table*}
\begin{tabular}{lcccccc}
\hline
Term & Frequency ($\Omega$) & $\sigma_\Omega$ & Amplitude ($A$) & $\sigma_A$ & Phase ($\mathbf{\phi}$) & $\sigma_\Phi$\\
\hline
\hline
1    & 2.01                 & 0.17     & 0.199         & 0.120    & 0.749          & 0.086    \\
2    & 6.23                 & 0.53     & 0.0724        & 0.3424   & 0.628          & 0.269    \\
3    & 5.23                 & 0.59     & 0.0428        & 0.3876   & 0.613          & 0.283    \\
4    & 1.06                 & 0.29     & 0.0465        & 0.1281   & 0.768          & 0.174    \\
5    & 3.97                 & 0.39     & 0.0676        & 0.2089   & 0.746          & 0.195    \\
6    & 6.99                 & 0.45     & 0.0587        & 0.3548   & 0.762          & 0.238    \\
7    & 7.89                 & 0.28     & 0.0394        & 0.1971   & 0.780          & 0.147   \\
\hline
\end{tabular}
\caption{Fourier fit parameters produced by \texttt{PERIOD04} for fitting to the \textit{y}-band light curve measured by \citet{popper57}. The fitting equation was $y=z+\sum A_i \sin(2\pi(\Omega_i t + \mathbf{\phi}_i))$ where $z=1.0870 \pm 0.017$. Errors were calculated using the Monte Carlo simulation mode in \texttt{PERIOD04}.}
\label{tbl:fitparams}
\end{table*}

\subsection{Polarization Position Angle}

The position angle (PA) of the intrinsic polarization diagnoses the dominant scattering angle and thus the orientation of the circumstellar material in the system. Panels (c) and (e) of Figure \ref{fig:pflux} show the PA variation with phase for the \textit{B} and \textit{V} bands of our V356 Sgr polarization data. Although the intrinsic PA is roughly constant with phase, stochastic variations of approximately $\pm5\degree$ in amplitude are clearly present. Such PA variations have been observed in the circumstellar environments of other disk-like systems \citep[e.g.,][]{carciofi2007,wis10,dra14}; these variations are usually interpreted as changes in the azimuthal density distribution in these disks. We return to this point in Section 4. 

Similar to our analysis of the system's optical broadband polarization behavior in Section 3.2, we computed the error-weighted mean position angle in the same four phase bins for all bands. We find that the average position angle is approximately constant ($\approx 75-76\degree$) across all phase bins (Table \ref{tbl:avgintrins}) and wavelengths. We subtracted the mean \textit{V}-band position angle (73.94\degree) from our dataset in \textit{q-u} space and display the residuals in Figure \ref{fig:sigma}. While the \textit{V}-band position angle displays more stochastic scatter during the first half of V356 Sgr's orbit than the second half, the fact that there are fewer observations at the later phases suggests that this may simply be an effect of uneven sampling. Therefore, we find no convincing trend in the stochastic position angle variations in our dataset as a function of phase. 

\subsection{UV Polarization}
Because the behavior of the ISP is less well understood in the UV, we removed the ISP from our WUPPE observation by extrapolating our optical ISP estimate (Section 3.2) to the UV wavelengths. \cite{Anderson} showed that this is not always a good assumption, but tends to estimate a lower bound on the UV ISP. In the worst-case scenario, we have undersubtracted the ISP by extending our optical ISP estimate into the UV.

\begin{figure}
\centering
\hspace*{-1.0cm}\includegraphics[width=0.52\textwidth]{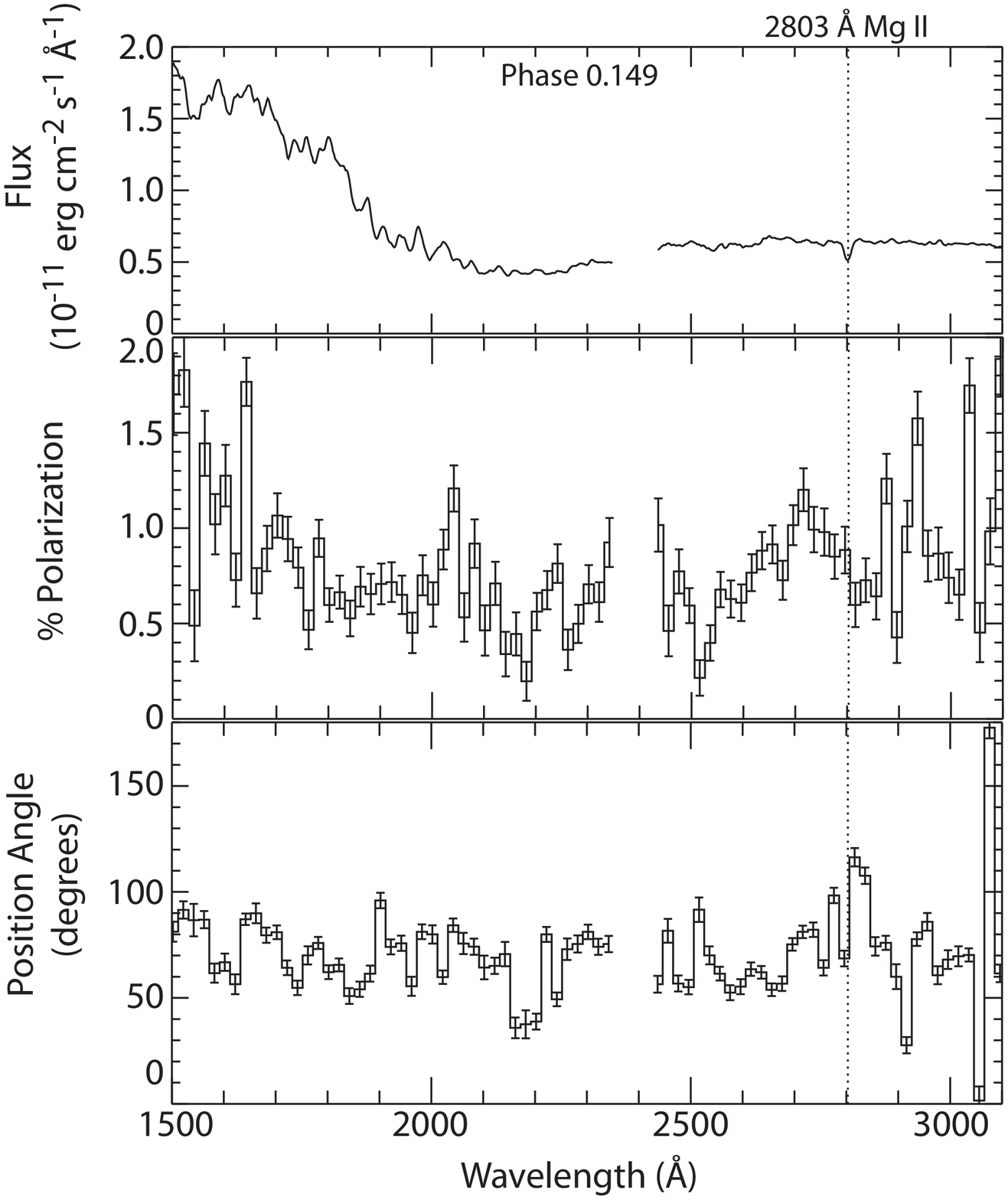}
\caption{Intrinsic UV polarization of V356 Sgr as observed by WUPPE at phase 0.149. The data are binned at 20\AA\space for display purposes. Data between 2348 \AA\space and 2436 \AA\space have been removed due to a known defect in the WUPPE detector affecting those wavelengths. \textit{Top panel:} UV spectrum of V356 Sgr. \textit{Middle panel:} Percent polarization as a function of wavelength. \textit{Bottom panel:} Position angle as a function of wavelength. V356 Sgr's UV polarization is relatively constant in the near UV. The only emission line that appears to deviate from the continuum polarization is the marked Mg II line. }
\label{fig:UVpol}
\end{figure}

We display our intrinsic UV polarimetric spectrum in Figure \ref{fig:UVpol}. We removed the data between 2348 \AA\space and 2436 \AA\space due to a known defect in the WUPPE detector that affects those wavelengths. The UV polarization appears higher at the extrema of this spectrum, but this may be due in part to the larger amount of scatter present in these regions. The position angle of the polarization is constant with wavelength. The error-weighted average (calculated in \textit{q-u} space) polarization and position angle ($0.641\%\pm0.013\%$ and $69.58\degree\pm0.58\degree$) of the system in the UV are similar to those of the \textit{B}-band optical data ($0.805\%$ and $76.15\degree$) displayed in Figure \ref{fig:pflux} and Table \ref{tbl:avgintrins}.

In contrast to the optical region, where there is no observable line polarization in our data, one UV absorption line appears to be polarized differently than the nearby continuum (Figure \ref{fig:UVpol}). The the position angle of the Mg II line at 2803 \AA\space appears similar to that of the continuum on the blue side but displays a different position angle than the continuum on the red side of the line. This behavior is independent of the bin size used to display the data. There also appears to be a decrease in the percent polarization across the line. However, both of these Mg II line features are similar in size to the size of the overall scatter in the polarization spectrum, making it difficult to determine their significance.

\section{Discussion: Scattering Regions \textbf{in V356 Sgr}}

Analyzing the phase variations of V356 Sgr's total brightness, intrinsic polarization magnitude, polarized flux, and position angle enables us to constrain the general geometry of the circumstellar material in the system. The lack of any significant change in the intrinsic polarization across primary eclipse indicates that when the donor passes in front of the gainer and its accretion disk, it does not substantially affect the amount of scattered light seen by the observer. This implies that the accretion disk around the gainer is not a significant scattering region, which suggests that there may be an additional region responsible for scattering photons within the system. The similarity between the morphology of V356 Sgr's polarized flux and total flux phase curves implies the same properties. The minima at primary and secondary eclipse in the polarized flux appear to be solely due to a decrease in the amount of total flux observed and not to any decrease in the percent polarization. Additionally, the position angle of V356 Sgr's polarization is roughly constant with phase. Taken together, these results suggest two possible scenarios for the system's scattering region(s); most of the visible light is scattered either by material out of the orbital plane of the binary or by circumbinary material in the orbital plane.

Regardless of the scattering scenario present in V356 Sgr, the data appear consistent with electron scattering; the intrinsic polarization is constant with wavelength. Therefore, no matter where the scattering region(s) in V356 Sgr are, they must be close enough to the stars to be ionized and produce a sea of electrons with which photons in the system can scatter. There is no indication of the presence of scattering with dust, which would have revealed itself through a wavelength dependent polarization signature.

In Be stars whose disk is denser than about $5\times 10^{ -12}$ g cm$^{ -3}$, the \ion{H}{1} opacity introduces a significant negative slope in the polarized spectrum (e.g., Fig. 1 of \citealt{Hau}). Therefore, the fact that the polarization of V356 Sgr is constant with wavelength implies that the amount of \ion{H}{1} opacity in the scattering region is not significant, and allows us to estimate an upper limit of roughly $5\times 10^{-12}$ g cm$^{-3}$ for the density of the scattering material.

\subsection{V356 Sgr's Accretion Disk}

We first consider the simplest possibility: that all the visible polarization we observe is produced within the known accretion disk surrounding the gainer. The fact that the intrinsic polarization does not vary substantially across primary eclipse suggests that this eclipse is an occultation of primarily unpolarized light. This in turn implies that the accretion disk surrounding the gainer is optically thin. In this scenario, light arising from the gainer scatters into our line of sight at the outer edges of the accretion disk, acquiring a small polarization magnitude and a position angle oriented perpendicular to the orbital plane; we can thus interpret the near-constant PA of the system as a measure of the orientation of the orbital plane on the sky. If the disk is symmetric, we can assume each of these scattering edges produces the same amount of polarization and the same PA. Thus, when the secondary star occults these edges in turn, the overall PA does not change. This is consistent with the PA phase curves in Figure 5. However, we would expect to see two sharp, repeatable dips in polarization, symmetric around the eclipse, as each scattering edge is occulted by the secondary star. This effect should also appear in the polarized flux curves, where it would be even more dramatic, because occultation of each edge would correspond to a loss of half the polarized flux out of eclipse. Neither effect appears in our data (Figure \ref{fig:pflux}). The fact that our phase curves are sparsely sampled in time may limit our ability to detect such a signal, and we intend to obtain more data at a higher sampling rate in the future to search for it. However, for purposes of this paper, we assume hereafter that the accretion disk in V356 Sgr is not the only source of polarization, and investigate two other possibilities to explain the polarization behavior we observe. Both may exist in combination with a small contribution from scattering in the accretion disk.

In Section 3.3, we noted that V356 Sgr exhibits stochastic variability in its intrinsic polarization at individual phases in all bandpasses (Figures \ref{fig:phasepolsub} and \ref{fig:sigma}).  Following \citet{wilson1995}, we posit that these variations may be due to small-scale changes in the mass transfer rate between the donor and gainer, which lead to small inhomogeneities in the density of scatterers in the gainer's accretion disk. This may change the amount of material in the accretion disk on small enough time scales to create the observed stochastic variability. For example, a large amount of material could be quickly injected into the accretion disk from the loser and then quickly accreted onto the gainer. This scenario likely holds even if the accretion disk is not the main source of polarization in the system, as discussed in the sections below.

\subsection{Extraplanar Material}

Light scattering in material outside of the orbital plane of the stars would explain the lack of observed eclipses in the polarization. If the extraplanar material is persistent and maintains a consistent geometry, it will imprint a PA on the observed polarization that is different from that produced by light scattering in the accretion disk. In this case, the observed near-constant PA does not reflect the orientation of the orbital plane on the sky. If scattering in the extraplanar material dominates the polarization, the observed PA traces the orientation of this material. A more likely scenario is that the observed PA represents an intermediate value produced by the combination of polarization from light scattering in the accretion disk and in the extraplanar material. If the extraplanar material is not persistent (i.e., exists sporadically or not in every orbit), then the average PA we observe may trace the orientation of the orbital plane, while the stochastic rotations away from this mean reflect transient concentrations of extraplanar material, perhaps related to variations in the accretion rate as discussed above. 

\cite{peters2004} found that the behavior of the system's UV emission lines across primary eclipse suggests that they form from material not near V356 Sgr's stars or their orbit, e.g., within bipolar outflows. This lends credence to the idea that there could be a region of scatterers outside of the plane of the orbit of the system. However, there is very little information about these bipolar outflows available in the literature; their extent above and below the plane of the disk is currently unknown and no other data set has confirmed their existence. 

If extraplanar material is the correct interpretation of V356 Sgr's polarimetric signature, it is worth noting that this material is quite different from the bipolar outflows in $\beta$ Lyr \citep{harmanec1996,hoffman1998}. In $\beta$ Lyr, bipolar outflows are responsible for scattering the optical and UV emission lines, as well as the UV continuum, while the disk scatters optical continuum light. The distinction between the two scattering regions manifests itself as a near $90\degree$ rotation in the position angle of the polarization across the lines and the Balmer jump. However, we do not observe this behavior in V356 Sgr. We detected no change in polarization across the optical lines in V356 Sgr, and we also found that the position angle of the UV polarimetric continuum is consistent (within $10\degree$) with that of its optical polarimetric continuum (Section 3.6). In this scenario, therefore, bipolar outflows or other extraplanar material would have to be responsible for the majority of the continuum polarization observed from V356 Sgr, without any significant contribution from V356 Sgr's disk.

\subsection{Circumbinary \textbf{or Intrabinary} Material}

Another possible interpretation of the observed polarimetric behavior of V356 Sgr is that it is due to scattering in circumbinary material in the system's orbital plane. Because this material is exterior to the orbit of the stars, very little of it is eclipsed at any given time. This explains the lack of observed eclipses and also the constancy of the observed position angle, which in this case would again trace the orientation of the orbital plane on the sky.

Circumstellar material between the stars but aligned with the orbital plane may also produce the polarization signatures we observe, as long as parts of the scattering region remain visible outside of both eclipses. One possibility for intrabinary material is the mass stream connecting the loser to the accretion disk; this is likely an elongated structure that lies in the orbital plane and is visible at most phases \citep{wilson1978}. Variations in the mass loss rate could cause the stream to become more or less dense over time and explain the stochastic changes in visible polarization we observe (Figure \ref{fig:pflux}). Another possible coplanar scattering region is the distorted atmosphere of the Roche-lobe-filling loser, a scenario proposed for $\beta$ Lyr by \citet{hoffman1998}, but this should show "eclipses" as the loser rotates, so by itself this cannot account for all the polarization we observe.

\subsection{Comparison to $\beta$ Lyrae}

Morphologically, $\beta$ Lyr and V356 Sgr are similar systems. They both are undergoing RLOF, which has caused disks to form around their gainers. Naturally, their similar geometry also causes both systems to have similarly shaped light curves. Polarimetrically speaking, the position angle of both systems is constant with phase and the magnitude of their polarization is similar (on the order of tenths of a percent). 
 
In a broader sense, we note that the observed polarimetric behavior of V356 Sgr is substantially different than that of the canonical eclipsing RLOF system $\beta$ Lyr \citep{hoffman1998,lomax2012}.  Whereas $\beta$ Lyr possesses clear intrinsic polarization changes across both primary and secondary eclipses, intra-eclipse polarimetric variations due to the orientations of the loser and disk, an optically thick accretion disk, and phase-dependent polarization in its emission lines, V356 Sgr exhibits none of these. We conclude that although both systems contain accretion disks and bipolar flows, the properties of their circumstellar material are quite different. First, as discussed above in this section, our results suggest that the accretion disk in V356 Sgr is not the dominant scattering region. Because of this, we are not able to detect a hotspot at the point where the mass stream impacts the accretion disk in V356 Sgr as was done in the case of $\beta$ Lyr \citep{lomax2012}. Thus, if V356 Sgr contains a hotspot, it must be detected through different means. In addition, while V356 Sgr may possess bipolar outflows \citep{peters2004}, they may not be evident in our spectropolarimetric data. In the case of $\beta$ Lyr, electron scattering in the outflows causes a PA rotation between the near-UV and visible continua and between the optical emission lines and continuum \citep{hoffman1998}. This is caused by UV photons being absorbed in the optically thick accretion disk in $\beta$ Lyr; the only way for UV photons to escape the system is to travel perpendicular to the accretion disk and scatter within $\beta$ Lyr's bipolar outflows. However, the accretion disk in the V356 Sgr system is optically thin (Section 4.1), while $\beta$ Lyr's disk is optically thick. Therefore, UV photons in V356 Sgr are able to travel to any scattering region in the system that optical photons also reach. Taken together, these results imply that the geometrical distribution of circumstellar material in RLOF systems exhibits more variety than previously understood. A comprehensive study of a larger number of such systems is needed to draw broad conclusions about the typical mode of mass transfer between massive stars in these systems.

\subsection{Limitations of the Data}
We searched our data set for trends, variations, and periodicities other than the orbital period of the system in an effort to better understand what may be causing the scatter seen in Figures \ref{fig:phasepolsub} and \ref{fig:pflux} and interpret the general behavior of our data. For example, we searched the data for long-term trends by binning the data on yearly time-scales, but found no obvious trends when comparing different years to each other. Similarly, we plotted the polarization and position angle versus the Julian date and found no trends spanning the time frame over which our data was taken; i.e. the measured polarization and position angle of V356 Sgr do not appear to be increasing or decreasing over time. For completeness we also used \texttt{PERIOD04} to search for important frequencies in addition to the orbital period of the system, but found nothing significant. 

Our efforts to characterize and interpret the orbital polarimetric variations of V356 Sgr and the scatter in our data set are hampered by a couple of factors. First, the data are not well sampled in time. V356 Sgr was observed in only seven of the 22 years that our data set spans. Most notably, there is a 13 year span, between 1999 and 2011, when the system was not monitored at all. Additionally, there is a wide range in the number of observations obtained per year during the seven years in which the system was monitored. The fewest observations, 2, occurred during 1994, while the most, 12, were obtained in 2012. Second, our data set does not adequately sample any single orbit of V356 Sgr. The combination of these two limitations make it difficult to assess the source of the scatter in our polarimetric observations. Using our existing data, we cannot rule out the possibility that the polarization of V356 Sgr is well-behaved over the course of one epoch, but exhibits large variations between cycles. Alternatively, it is possible that the system does not display repeatable periodic behavior on any timescale. Future polarimetric observing campaigns could better investigate the cause of observed variations and scatter by intensely focusing on V356 Sgr over the course of one or two epochs (e.g. with observations at least once or twice per night) and by continuing to monitor the system over several more cycles at a lower frequency (e.g. one observation every 2-3 nights for several weeks).  

Additionally, there is currently very little published information about the UV spectrum and polarization of V356 Sgr. UV spectra taken during the system's primary eclipse have been published \citep{peters2004}, but there are no data that cover the rest of V356 Sgr's orbital period. There also exists two UV polarimetric observations in the HST archive that have yet to appear in the literature outside of conference abstracts (e.g. \citealt{polidan}). While beyond the scope of this paper, we plan to analyze and compare these data to our WUPPE observation in the near future in order to better understand the UV polarimetric behavior of the system.

Finally, our ISP estimate is not rigorous. As previously stated, changes in our estimate of the ISP's position angle by $10\degree$ did not significantly affect our results. Perhaps even more problematic is that the field stars we used are quite far apart from both each other and V356 Sgr. Because of this, we may be probing different ISM clouds with different field stars which may in turn lead to a poor understanding of how the ISM is affecting the observed V356 Sgr polarization. Future polarimetric work on this system might benefit from identifying an ISP probe star that is spatially close to V356 Sgr with a small angular separation. However this may prove to be difficult; V356 Sgr does not lie within a cluster of stars.

\section{Summary and Future Prospects}
We have presented the first comprehensive optical spectropolarimetric study of the eclipsing RLOF system V356 Sgr. After characterizing and removing the interstellar polarization contribution to our observations, we find V356 Sgr exhibits $\sim$1\% intrinsic polarization that displays no significant wavelength dependence. This is consistent with electron scattering in material within the V356 Sgr system. Our results suggest that the known accretion disk in the V356 Sgr system is optically thin and likely not the dominant source of scattered light in the system. The flat spectrum, furthermore, allows us to rule out any significant dust scattering and \ion{H}{1} absorption. We estimate that the upper limit for the density of the scattering material is $5\times 10^{-12}$ g cm$^{ -3}$. We interpret the lack of significant intrinsic polarization variations across primary and secondary eclipses as evidence of a substantial body of scatterers outside the stars and disk in the system, located either in the orbital plane (circumbinary material or a mass stream) or external to the plane (possible bipolar outflows). We also observed a large amount of stochastic polarimetric variability, which we suggest indicates the mass transfer varies on short timescales, causing rapid variability in the density of scatterers present within the system's accretion disk. We also note multiple fundamental differences between V356 Sgr and the canonical RLOF system, $\beta$ Lyr, which suggest there is greater diversity in the basic circumstellar geometry of RLOF systems than previously established. 

In order to better understand the V356 Sgr system, we suggest that future polarimetric observations of the system be carried out in a dedicated and consistent manner. Obtaining data multiple times a night, every night, over two or more orbital periods will likely reveal the cause of the stochastic variability we detected here. Additionally, longer time-scale monitoring conducted in a consistent manner may be able to clearly reveal whether the system is changing with time. Therefore, we plan to obtain more polarimetric data of V356 Sgr using this observing strategy in the near future. Analysis of the existing archival V356 Sgr UV polarimetric observations is needed, as well as new UV observations of the system. Future interferometric observations of V356 Sgr will be able to distinguish between the scattering geometry scenarios we present in Section 4 by definitively determining the position angle of the orbit of the binary on the sky.

\section*{Acknowledgments}
We gratefully acknowledge the contributions of the large team of HPOL and WUPPE observers from the University of Wisconsin who helped to acquire and reduce these data. The relocation and refurbishment of HPOL was supported in part by a Small Research Grant from the American Astronomical Society and by funding from the Dunham Fund for Astrophysical Research. Observations with HPOL at Ritter Observatory have been supported by NSF grant AST-1412135 (KSB/JEB), the Helen Luedtke Brooks Endowed Professorship (KSB), and the Scott Smith Fund for Research at Ritter Observatory. We thank the Ritter observing team for help with acquiring observations. We also thank G. Peters and A.M. Magalh\~{a}es for helpful discussions in the early stages of this project. A.C.C. acknowledges support from CNPq (grant 307594/2015-7) and FAPESP (grant 2015/17967-7). This research has made use of the SIMBAD database, operated at CDS, Strasbourg, France. Some of the data presented in this paper were obtained from the Mikulski Archive for Space Telescopes (MAST). STScI is operated by the Association of Universities for Research in Astronomy, Inc., under NASA contract NAS5-26555. Support for MAST for non-HST data is provided by the NASA Office of Space Science via grant NNX09AF08G and by other grants and contracts.
\nocite{*}

\begin{thebibliography}{9}
\bibitem[Anderson et al.(1996)]{Anderson} Anderson, C.~M., Weitenbeck, A.~J., Code, A.~D., et al.\ 1996, \aj, 112, 2726 

\bibitem[Bessell(1990)]{bessell1990} Bessell, M.~S. \ 1990, \pasp, 102, 1181

\bibitem[Bessell(2005)]{Bessell2005} Bessell, M.~S.\ 2005, \araa, 43, 293

\bibitem[Bjorkman et al.(1993)]{Bjorkman} Bjorkman, K.~S., Meade, M.~R., Nordsieck, K.~H., et al.\ 1993, \apj, 412, 810

\bibitem[Brown et al.(1978)]{brown1978} Brown, J.~C., McLean, I.~S., \& Emslie, A.~G.\ 1978, \aap, 68, 415 

\bibitem[{Carciofi} et al.(2007)]{carciofi2007} {Carciofi}, A.~C., {Magalh{\~a}es}, A.~M., {Leister}, N.~V. et al.\ 2007, \apj, 671, L49

\bibitem[Clarke \& Bjorkman(1998)]{cla98} Clarke, D. \& Bjorkman, K.S. 1998, \aap, 331, 1059

\bibitem[Davidson et al.(2014)]{davidson2014} Davidson, J.~W., Bjorkman, K.~S., Hoffman, J.~L., et al.\ 2014, J. of  Astron. Inst. , 3, 1450009

\bibitem[de Mink et al.(2007)]{deMink2007} de Mink, S.~E., Pols, O.~R., \& Hilditch, R.~W.\ 2007, \aap, 467, 1181 
    
\bibitem[Draper et al.(2014)]{dra14} Draper, Z.H., Wisniewski, J.P., Bjorkman, K.S., Meade, M.R., Haubois, X., Mota, B.C., Carciofi, A.C., \& Bjorkman, J.E. 2014, ApJ, 786, 120

\bibitem[Eggen, Gascoigne, \& Burr(1957)]{egg57} Eggen, O.J., Gascoigne, S.C.B., \& Burr, E.J. 1957, \mnras, 117, 406

\bibitem[Haubois et al.(2014)]{Hau} Haubois, X., Mota, B.~C., Carciofi, A.~C., et al.\ 2014, \apj, 785, 12

\bibitem[Hall et al.(1981)]{hall1981} Hall, D.~S., Henry, G.~W., \& Murray, W.~H.\ 1981, \actaa, 31, 383

\bibitem[Hall(1958)]{Hall1958} Hall, J.~S.\ 1958, Publications of the U.S.~Naval Observatory Second Series, 17, 275

\bibitem[Harmanec et al.(1996)]{harmanec1996} Harmanec, P., Morand, F., Bonneau, D., Jiang, Y., Yang, S., Guinan, E.~F., Hall, D.~S., Mourard, D., Hadrava, P., Bozic, H., Sterken, C., Tallon-Bosc, I., Walker, G.~A.~H., McCook, G.~P., Vakili, F., Stee, P., Le Contel, J.~M. \ 1996, \aap, 312, 879

\bibitem[Heiles(2000)]{heiles2000} Heiles, C.\ 2000, \aj, 119, 923 

\bibitem[Hoffman et al.(1998)]{hoffman1998} Hoffman, J.~L., Nordsieck, K.~H., \& Fox, G.~K.\ 1998, \aj, 115, 1576



\bibitem[Iben(1991)]{Iben1991} Iben, I., Jr.\ 1991, \apjs, 76, 55 


\bibitem[Lenz \& Breger(2005)]{lenz2005} Lenz, P., \& Breger, M.\ 2005, Communications in Asteroseismology, 146, 53 

\bibitem[Lomax et al.(2012)]{lomax2012} Lomax, J.~R., Hoffman, J.~L., Elias, N.~M., II, Bastien, F.~A., \& Holenstein, B.~D.\ 2012, \apj, 750, 59

\bibitem[Magalh\~{a}es et al.(1996)]{magalhaes1996} {Magalh\~{a}es}, A.~M., {Rodrigues}, C.~V., {Margoniner}, V.~E. et al.\ 1996, Astronomical Society of the Pacific Conference Series, 97, 118

\bibitem[Mathewson \& Ford(1970)]{mat70} Mathewson, D.S. \& Ford, V.L. 1970, MNRAS, 74, 139

\bibitem[McLean \& Clarke(1979)]{mcl79} McLean, I.S. \& Clarke, D. 1979, MNRAS, 186, 245

\bibitem[Nook(1990)]{nook1990} Nook, M.A. 1990, Ph.D. thesis, University of Wisconsin-Madison

\bibitem[Nordsieck et al.(1994)]{wuppe} Nordsieck, K.~H., Code, A.~D., Anderson, C.~M., et al.\ 1994, \procspie, 2010, 2 

\bibitem[Nordsieck \& Harris(1996)]{nordsieck1996} Nordsieck, K.~H., \& Harris, W.\ 1996, Polarimetry of the Interstellar Medium, 97, 100 

\bibitem[Peters \& Polidan(2004)]{peters2004} Peters, G.~J., \& Polidan, R.~S.\ 2004, Astronomische Nachrichten, 325, 225 


\bibitem[Podsiadlowski et al.(1992)]{Podsiadlowski1992} Podsiadlowski, P., Joss, P.~C., \& Hsu, J.~J.~L.\ 1992, \apj, 391, 246 


\bibitem[Polidan et al.(1994)]{polidan} Polidan, R.~S., Lynch, D.~E., Downes, R.~A., Keyes, C.~D., \& Shore, S.~N.\ 1994, baas, 26, 85.23 

\bibitem[Polidan \& Lynch(1996)]{Polidan96} Polidan, R.~S., \& Lynch, D.~E.\ 1996, baas, 28, 38.04 

\bibitem[Popper(1955)]{popper1955} Popper, D.~M.\ 1955, \apj, 121, 56 

\bibitem[Popper(1957)]{popper57} Popper, D.~M.\ 1957, \apjs, 3, 107 

\bibitem[Popper(1980)]{popper1980} Popper, D.~M.\ 1980, \araa, 18, 115 

\bibitem[Quirrenbach et al.(1997)]{qui97} Quirrenbach, A., Bjorkman, K.S., Bjorkman, J.E., Hummel, C.A., Buscher, D.F., Armstrong, J.T., Mozurkewich, D., Elias II, N.M., \& Babler, B.L. 1997, ApJ, 479, 477

\bibitem[Sana et al.(2012)]{sana2012} Sana, H., de Mink, S.~E., de Koter, A., et al.\ 2012, Science, 337, 444 

\bibitem[Serkowski et al.(1975)]{serkowski1975} Serkowski, K., Mathewson, D.~S., \& Ford, V.~L.\ 1975, \apj, 196, 261 

\bibitem[Smartt(2009)]{Smartt2009} Smartt, S.~J.\ 2009, \araa, 47, 63 


\bibitem[van Leeuwen(2007)]{vanleeuwen2007} van Leeuwen, F.\ 2007, \aap, 474, 653 

\bibitem[van Rensbergen et al.(2011)]{rensbergen2011} van Rensbergen, W., de Greve, J.~P., Mennekens, N., Jansen, K., \& de Loore, C.\     2011, \aap, 528, AA16 

\bibitem[Wenger et al.(2000)]{SIMBAD} Wenger, M., Ochsenbein, F., Egret, D., et al.\ 2000, \aaps, 143, 9 
    
\bibitem[Wilking et al.(1982)]{wil82} Wilking, B. A., Lebofsky, M. J., \& Rieke, G. H. 1982,
AJ, 87, 695    


\bibitem[Wilson \& Caldwell(1978)]{wilson1978} Wilson, R.~E., \& Caldwell, C.~N.\ 1978, \apj, 221, 917

\bibitem[Wilson \& Woodward(1995)]{wilson1995} Wilson, R.~E., \& Woodward, E.~J.\ 1995, \pasp, 107, 132 

\bibitem[Wisniewski et al.(2007)]{wis07} Wisniewski, J.P., Bjorkman, K.S., Magalhaes, A.M., Bjorkman, J.E., Meade, M.R., Pererya, A. 2007, 
ApJ, 671, 2040

\bibitem[Wisniewski et al.(2010)]{wis10} Wisniewski, J.P., Draper, Z.H., Bjorkman, K.S., Meade, M.R., Bjorkman, J.E., \& Kowalski, A.F. 2010, ApJ, 709, 1306

\bibitem[Wolff et al.(1996)]{wolff1996} Wolff, M.~J., Nordsieck, K.~H., \& Nook, M.~A.\ 1996, \aj, 111, 856 

\bibitem[Zhao et al.(2008)]{zhao2008} Zhao, M., Gies, D., Monnier, J.~D., et al.\ 2008, \apjl, 684, L95 

\bibitem[Ziolkowski(1985)]{ziolkowski1985} Ziolkowski, J.\ 1985, \actaa, 35, 199
\end{thebibliography}



\label{lastpage}

\end{document}